\documentstyle[12pt]{article}
\begin{document}\openup6pt

\title{\bf Radiating spherical collapse with heat flow}

\author{
{\bf M Govender\thanks{megang@ntech.ac.za}}\\
Department of Physics, Technikon Natal, PO Box 953\\
Durban 4000, South Africa\\[2mm]
{\bf K S Govinder\thanks{govinder@nu.ac.za}, S D 
Maharaj\thanks{maharaj@nu.ac.za}}\\
School of Mathematical and Statistical Sciences\\
University of Natal, Durban 4041, South Africa\\[2mm]
{\bf R Sharma\thanks{ranjan$\_$sh@hotmail.com}, S 
Mukherjee\thanks{sailom47@hotmail.com}}\\
Department of Physics,
North Bengal University\\
Dist. Darjeeling 734 430, India\\[2mm]
and\\[2mm]
{\bf T K Dey\thanks{tkdey@dte.vsnl.net.in}}\\
Physics Department, Gurucharan College\\
Silchar 788 004, India\\
}

\date{}

\maketitle

\begin{abstract}
We present here a simple model of radiative gravitational collapse with radial
heat flux which describes qualitatively the stages close to the formation of a superdense cold star. Starting with a static general solution for a cold star, 
the
model can generate solutions for the earlier evolutionary stages.
The temporal evolution of the model is
specified by solving the junction conditions appropriate for radiating
gravitational collapse.

\end{abstract}

\newpage

\section{INTRODUCTION}

The theory of gravitional collapse has many interesting applications in
astrophysics where the formation of cold compact stellar objects are usually preceded  by a period of radiative collapse.
The problem of gravitional collapse was first investigated by Oppenheimer 
and
Snyder~\cite{kn:1}, who considered the contraction of a spherically
symmetric dust cloud. Here the exterior spacetime was described  by the
Schwarzschild metric and the interior space time was represented by a 
Friedman-like solution. Thereafter, Vaidya~\cite{kn:2} derived the line element which 
describes the
exterior gravitational field of a spherically symmetric radiating mass.
It then became possible to model the interior of radiating stars by 
matching
the interior solution to the exterior space time given by Vaidya~\cite{kn:3,kn:4,kn:5,kn:6,kn:7}. The junction conditions for a
spherically symmetric shear-free radiating star was completely derived by Santos~\cite{kn:8}.
The crucial result that follows from Santos is that the pressure on the
boundary of a radiating sphere is nonvanishing in general. Subsequently,
many models of radiative gravitational collapse with heat flow were found
by utilising these junction conditions. (See, e.g. \cite{kn:16}.) 
In particular, special attention was
given to models in which an initial static stellar configuration began
collapsing by dissipating energy in the form of a radial heat flux~\cite{kn:9}. The initial static configuration was taken to be an exact
solution of the Einstein field equations. 

The aim of this paper is to consider the evolution of a star undergoing 
radiative gravitional collapse with its {\it final} state being that of a 
superdense
star. We wish to study this evolutionary process starting from the final 
nonradiating
state and interpolating to earlier times when it was emitting radiation.

In Section 2 we present the relevant background material and the line 
element
for the interior space time. The Einstein equations for an energy momentum
tensor with heat flux are solved without assuming a particular form for the
final static configuration. The Vaidya solution is introduced in Section 3
and the matching condition is utilised to determine approximately the 
temporal evolution of the model. In Section 4 we use the solution 
of Mukherjee {\em et al}~\cite{kn:10} to describe the final state of the star and
find out its evolution by making use of the final 
static configuration. Although we are using a simple model, the results obtained analytically are expected to provide the general trends of results for realistic stars. In Section 5, we summarize
our results.

\section{INTERIOR SPACETIME}

Let us assume that the interior spacetime of a star is represented by a line element of the form
\begin{flushleft}
\hspace{1in} $ds^2 = -A^2(t,r)dt^2 + B^2(t,r)(dr^2 + r^2d\Omega^2)$
\hfill(1)
\end{flushleft}
The choice of a shear-free metric is motivated by the simplicity of the resulting equations, which remain tractable throughout. Moreover, as observed by Bonnor {\em et al}~\cite{kn:9}, it is possible to show through Raychaudhuri's equation that the slowest possible collapse is for shear-free fluids. Thus our results will be appropriate in any case if the collapse is not very fast. Also, the choice of a shear-free metric will later on help us to apply the same junction conditions derived by Santos~\cite{kn:8} in our model. 

The energy momentum tensor for the interior matter distribution is taken to be
\begin{flushleft}
\hspace{1in} $T_{\mu\nu} = (\rho + p)u_\mu u_\nu + pg_{\mu\nu} +q_\mu u_\nu
         + q_\nu u_\mu$ \hfill(2)
\end{flushleft}
where the flow vector satisfies
$$q^\mu u_\nu = 0,$$
$u^\mu$ being a timelike four velocity vector. 

Following Bonnor {\em et al}~\cite{kn:9}, we choose the metric functions as follows:
\begin{flushleft}
\hspace{1in} $A(r,t) = A_0(r) + \epsilon a(r) T(t)$ \hfill(3)

\hspace{1in} $B(r,t) = B_0(r) + \epsilon b(r) T(t)$ \hfill(4)
\end{flushleft}

and the energy density $\rho$ and the isotropic pressure $p$ as
\begin{flushleft}
\hspace{1in} $\rho(r,t) = \rho_0(r) + \epsilon\bar{\rho}(r,t)$
\hfill(5)

\hspace{1in} $p(r,t) = p_0(r) + \epsilon\bar{p}(r,t)$
\hfill(6)
\end{flushleft}

The radial heat flux is of the order of $\epsilon$ $(0 < \epsilon \ll 1)$.
However, unlike the case studied by Bonner {\em et al}~\cite{kn:9}, $A_0$ and $B_0$ 
describe here the final static solutions of the cold star.

Einstein's field equations for the static configuration give the relations:
\begin{flushleft}
\hspace{1in} $\rho_0 = 
-\frac{1}{B_0^2}\left[2\left(\frac{B_0''}{B_0}\right)-
               \left(\frac{B_0^\prime}{B_0}\right)^2 + \frac{4}{r}\left(
               \frac{B_0^\prime}{B_0}\right) \right]$ \hfill(7)
\end{flushleft}
\begin{flushleft}
\hspace{1in} $p_0 = \frac{1}{B_0^2} \left[ 
\left(\frac{B_0^\prime}{B_0}\right)^2 +
            2\frac{A_0^\prime}{A_0}
           \frac{B_0^\prime}{B_0} + \frac{2}{r}\left(\frac{A_0^\prime}{A_0}+
            \frac{B_0'}{B_0}\right)
              \right ]$ \hfill(8)
\end{flushleft}
where prime $(\prime)$ denotes differentiation with respect to $r$.

The pressure isotropy equation is given by
\begin{flushleft}
\hspace{1in}
$\left(\frac{A_0^\prime}{A_0} + \frac{B_0^\prime}{B_0} \right)^\prime -
\left(\frac{A_0^\prime}{A_0} + \frac{B_0^\prime}{B_0} \right)^2 -
\frac{1}{r}
  \left(\frac{A_0^\prime}{A_0} + \frac{B_0^\prime}{B_0} \right) +
  2\left( \frac{A_0^\prime}{A_0}\right)^2 = 0$ \hfill(9)
\end{flushleft}

We assume that the presure isotropy condition is satisfied for a known static 
solution
$(A_0,B_0)$. The perturbed field equations up to first order in $\epsilon$ 
can be written as 
\begin{flushleft}
\hspace{1in}
$\bar{\rho} = - 3 \rho_0 \frac{b}{B_0} T + \frac{1}{B_0^3} \left[
              -\left( \frac{B_0'}{B_0}\right)^2 b + 2\left(\frac{B_0'}{B_0}
                -\frac{2}{r}\right) b' - 2 b'' \right] T$ \hfill(10)

\hspace{.5in}
$\bar{p} = - 2 p_0 \frac{b}{B_0} T + \frac{2}{B_0^2} \left[ \left(\frac{
                B_0'}{B_0} + \frac{1}{r} + \frac{A_0'}{A_0} \right)
                \left(\frac{b}{B_0}\right)' +
                 \left(\frac{B_0'}{B_0} + \frac{1}{r}\right) \left(
                 \frac{a}{A_0}\right)' \right] T - 2 \frac{b}{A_0^2 B_0}
                \ddot T$ \hfill(11)

\hspace{1in}
$q = \frac{2\epsilon}{B_0^2} \left( \frac{b}{A_0 B_0}\right)' \dot T $\hfill(12)
\end{flushleft}
where an overhead dot denotes differentiation with respect to $t$.

The condition of pressure isotropy for the perturbed matter 
distribution can be written as
\begin{flushleft}
\hspace{.1in}
$\left[ \left(\frac{a}{A_0}\right)' + \left(\frac{b}{B_0}\right)' \right]'
  - 2 \left[ \left(\frac{a}{A_0}\right)' + \left(\frac{b}{B_0}\right)' 
\right]
    \left(\frac{A_0'}{A_0} + \frac{B_0'}{B_0}\right) - $ \\ \hspace{2in} $
    \frac{1}{r} \left[
     \left(\frac{a}{A_0}\right)' + \left(\frac{b}{B_0}\right)' \right]
    + 4 \frac{A_0'}{A_0}\left(\frac{a}{A_0}\right)' = 0$ \hfill(13)
\end{flushleft}

The general solution to this equation is
\begin{flushleft}
\hspace{.1in}
$\frac{b}{B_0} = C_0\int r A_0^2 B_0^2 dr + C_1 - \int r A_0^2 B_0^2 \int 
A_0^{-4}
\left(\frac{(\frac{a}{A_0})' A_0^2}{r B_0^2}\right)' dr dr $ \hfill(14)
\end{flushleft}
or, equivalently,
\begin{flushleft}
\hspace{.1in}
$\frac{a}{A_0} = C_0 \int \frac{r B_0^2}{A_0^2} dr +C_1 - \int \frac{r 
B_0^2}{A_0^2}
\int A_0^4 \left(\frac{(\frac{b}{B_0})'}{r A_0^2 B_0^2}\right)' dr dr $ 
\hfill(15)
\end{flushleft}
where $C_0$ and $C_1$ are constants.

Equation (14) or (15) gives a relation between $a(r)$ and $b(r)$. For 
specific calculations, however, it
will be helpful to follow an alternative approach. We write 
$\left(\frac{a}{A_0}\right)'= X(r)$ and $\left(\frac{b}{B_0}\right)'= Y(r)$ and 
rewrite equation (13) as
\begin{flushleft}
\hspace{1in}
$\frac{(X + Y)'}{X + Y} - \left( \frac{2A_0'}{A_0} + \frac{2B_0'}{B_0}
     + \frac{1}{r} \right) + 4 \frac{A_0'}{A_0} \frac{X}{X+Y} = 0 .$
   \hfill(16)
\end{flushleft}

Equation (16) can be integrated easily if we assume
\begin{flushleft}
\hspace{1in}
$\frac{X}{X + Y} = A_0 \frac{dg(A_0)}{dA_0} $ \hfill(17)
\end{flushleft}
which gives
\begin{flushleft}
\hspace{1in}
$X = k_1 r A_0^3 B_0^2 y e^{-4 g(A_0)}$\hfill(18)
\end{flushleft}
and
\begin{flushleft}
\hspace{1in}
$Y = \left(\frac{1}{yA_0}-1\right) X $\hfill(19)
\end{flushleft}
where $k_1$ is an integration constant and $y = \frac{dg(A_0)}{dA_0}$.

Thus we obtain
\begin{flushleft}
\hspace{1in}
$\frac{a}{A_0} = k_1\int r A_0^3 B_0^2 y e^{-4 g(A_0)} dr +k_2 $ \hfill(20)
\end{flushleft}
where $k_2$ is another integration constant.

The right hand side of equation (20) can be integrated for different choices
of the function $g(A_0)$. It is instructive to consider a simple case:
\begin{flushleft}
\hspace{1in}
$g(A_0) = \frac{1}{2}\ln A_0 $\hfill(21)
\end{flushleft}

This gives
\begin{flushleft}
\hspace{1in}
$\frac{a}{A_0} = \frac{b}{B_0} = \frac{k_1}{2}\int r B_0^2 dr+ k_2 $. 
\hfill(22)
\end{flushleft}

It may be useful to calculate the total energy entrapped
within the surface $\Sigma$ of the star. Up to first order in $\epsilon$, 
this is given by
\begin{flushleft}
\hspace{1in}
$m(r_{\Sigma},t) = m_0(r_{\Sigma}) + \epsilon \bar m(r_{\Sigma},t)$ 
\hfill(23)
\end{flushleft}
where
\begin{flushleft}
\hspace{1in}
$m_0(r_{\Sigma}) = -\left(r^2 B_0' + r^3 \frac{B_0'^2}{2 
B_0}\right)_{\Sigma}$ \hfill(24)
\end{flushleft}

\begin{flushleft}
\hspace{1in}
$\bar m(r_{\Sigma},t) =\left(\left[-r^2 b' - r^3 \frac{B_0'^2}{2 B_0} 
\left(2\frac{b'}{B_0'}-\frac{b}{B_0}\right)\right] T(t)\right)_{\Sigma}$
\hfill(25)
\end{flushleft}
where $r_{\Sigma}$ corresponds to the boundary. The evaluation of 
$m_0(r_{\Sigma})$ and $\bar m (r_{\Sigma},t)$ will
be pursued in Section 4.

\section{JUNCTION CONDITIONS}

The boundary of the collapsing star divides spacetime into two distinct
regions, the interior spacetime described by the metric (1) and the exterior spacetime. 
Since the collapsing star is radiating energy, the exterior spacetime is described 
by Vaidya's outgoing metric
\begin{flushleft}
\hspace{1in}
$ds^2 = - \left(1 - \frac{2m(v)}{r}\right) dv^2 - 2dv dr + r^2( d\theta^2
       + \sin^2\theta d\phi^2)$ \hfill(26)
\end{flushleft}
$m(v)$ being an arbitrary function of the retarded time $v$. The solution 
(26) is the unique spherically symmetric solution of the Einstein field equations 
for radiation in the form of a null fluid. The Vaidya solution is often used to 
describe the exterior gravitational field of a radiating star~\cite{kn:3,kn:4,kn:5,kn:6,kn:11}.

In order to facilitate the smooth matching of the interior space-time 
to the Vaidya exterior, we utilise the junction conditions deduced by Santos~\cite{kn:8}. We can rewrite equations (11) and (12) as
\begin{flushleft}
\hspace{1in}
$\bar p = - 2p_0\frac{b}{B_0} T + 2 \frac{b}{A_0^2 B_0}\left(\alpha T-\ddot 
T\right)$ \hfill(27)
\end{flushleft}

\begin{flushleft}
\hspace{1in}
$q= \frac{4b\epsilon}{A_0^2 B_0^2}\beta \dot T$ \hfill(28)
\end{flushleft}
where
\begin{flushleft}
\hspace{1in}
$\alpha= \frac{A_0^2}{b B_0}\left[\left(\frac{B_0'}{B_0}+\frac{1}{r}+
\frac{A_0'}{A_0}\right)
\left(\frac{b}{B_0}\right)' + \left(\frac{B_0'}{B_0}+\frac{1}{r}\right)
\left(\frac{a}{A_0}\right)'\right]$ \hfill(29)
\end{flushleft}

\begin{flushleft}
\hspace{1in}
$\beta = \frac{A_0^2}{2b}\left(\frac{b}{A_0 B_0}\right)'$. \hfill(30)
\end{flushleft}

To find $T(t)$ we make use of the junction condition
$p_{\Sigma}= \left(q B\right)_{\Sigma}$ together with
$\left(p_0\right)_{\Sigma} = 0$ which gives
\begin{flushleft}
\hspace{1in}
$\alpha_{\Sigma} T -\ddot T = 2\beta_{\Sigma} \dot T$ \hfill(31)
\end{flushleft}
where $\Sigma$ represents the boundary of the star. The solution of (31) is 
given by

\begin{flushleft}
\hspace{1in}
$T(t) = T_0 \exp\left[-\left(\beta_{\Sigma} +
\sqrt{\alpha_{\Sigma}+{\beta_{\Sigma}}^2}\right) t\right]$ \hfill(32)
\end{flushleft}
which satisfies the boundary conditions

\begin{flushleft}
\hspace{1in}
$T(t)|_{t=\infty} = 0$ and $T(t)|_{t=0} = T_0$ \hfill(33)
\end{flushleft}
where $T_0$ is a constant. Since we expect $T(t)$ to decrease as $t$ increases, we have $\alpha_{\Sigma} > 0$. (see later.)

\section{FINAL STATIC SOLUTION}

We assume that the line element (1) represents a static solution at large 
times i.e. as $t \rightarrow \infty$, and the metric coefficients are then
represented by $A_0$ and $B_0$, the static part of equations (3) and (4), 
respectively.

For the static solution we take the solutions of Mukherjee {\em et al}~\cite{kn:10} of the Vaidya-Tikekar~\cite{kn:17} model for a class of static spherically symmetric stars, viz.

\begin{flushleft}
\hspace{1in}
$ds^2 = -e^{2 \nu(\bar r)} dt^2 + e^{2\mu(\bar r)} d\bar r^2 + \bar 
r^2(d\theta^2 +
\sin^2 \theta  d \phi^2)$ \hfill(34)
\end{flushleft}
where

\begin{flushleft}
\hspace{1in}
$e^{2\mu(\bar r)}= \frac{1+\lambda\bar r^2/R^2}{1-\bar r^2/R^2}$ \hfill(35)
\end{flushleft}
and

\begin{flushleft}
\hspace{1in}
$e^{\nu(\bar r)} = D\bigg[{\cos[(n+1)\zeta +\omega]\over n+1}-
{\cos[(n-1)\zeta
+\omega]\over n-1}\bigg]$ \hfill(36)
\end{flushleft}

In equations (35) and (36), $D$, $\omega$ and $R$ are constants, $ \zeta =\cos^{-1} z$, 
$n^2=\lambda+2 $ and
$z^2=(\frac{\lambda}{\lambda+1})(1-\frac{\bar r^2}{R^2})$.

The solution satisfies strong and weak energy conditions as well as the  causality condition and is valid for any value of $\lambda \geq \frac{3}{17}$~\cite{kn:10}. The boundary of the star $b$ is determined by imposing the condition $p_{0} = 0$ at $\bar r = b$. At the boundary of the star, the solution is matched to the Schwarzchild exterior solution. The model has four parameters $\lambda$, $R$, $D$ and $\omega$, of which three, say $R$, $D$ and $\omega$, are determined by the matching condition ($p_{0} = 0$ at $\bar r = b$ and the continuity of two metric coefficients). The choice of the remaining parameter $\lambda$, then, fixes the relevant equation of state 
of the star. Thus, in the present model, for a given mass and radius, there is a one-parameter class of equations of state parametrized by $\lambda$. Previous works \cite{kn:12,kn:14} show
that this model can describe cold compact stars like Her X-1 and SAX J1808.4-3658. The model also has a scaling property which allows a description of a class of stars with the same compactness~\cite{kn:18}. Moreover, the stability of the stellar configurations for various $\lambda$ can be verified easily in this model, as shown in \cite{kn:14}. Thus although the solution provides a toy model of a star, its simple analytic form is found to be very useful in studying the gross features of a star with an equation of state specified by the value of $\lambda$.

However, to make use of this solution, we need to transform 
the metric (1)
to the form given by (34), i.e. we need to change the radial coordinate $r$ to $\bar r$.

A comparison of equations (1) and (34) yields
\begin{flushleft}
\hspace{1in}
$\bar r = r B_0(r)$ and $e^{\mu(\bar r)} d\bar r = B_0(r) dr$ \hfill(37)
\end{flushleft}

Using equations (35) and (37), we obtain the inverse transformation relation as
\begin{flushleft}
\hspace{1in}
$r = k_3 \exp\left[ -\tanh^{-1}\left(\frac{\sin{\chi}}{\sqrt{1+\lambda 
\cos^2{\chi}}}\right) -
\sqrt{\lambda} 
\sin^{-1}\left(\sqrt{\frac{\lambda}{\lambda+1}}\sin{\chi}\right)\right]$ 
\hfill(38)
\end{flushleft}
where $\chi = \cos^{-1} \left(\frac{\bar r}{R}\right)$ and $k_3$ is an 
integration constant.

The integration in equation (22) is over $r$, but using equation (37) we can
integrate it over $\bar r$ so that
the final results are expressed in terms of $\bar r$. We obtain

\begin{flushleft}
\hspace{1in}
$\frac{a}{A_0} = \frac{b}{B_0} = - \frac{k_1}{4 l^2}R^2 \sqrt{\lambda}\left[\sin^{-1} (l 
x) + l x \sqrt{1-l^2 x^2}\right] +k_2$ \hfill(39)
\end{flushleft}
where $l=\sqrt{\frac{\lambda}{\lambda+1}}$, $ x=\sqrt{1-\frac{\bar 
r^2}{R^2}}$.

To get a feel for the behaviour of the solution, we need to consider specific numeric cases.
As an example, we consider a star whose final state has a mass $m=0.88 M_{\odot}$ and radius $b=7.7$ km. Note that, these values fall well within the estimated mass and radius of the well known X-ray pulsar Her X-1~\cite{kn:19}. Moreover, with a particular choice 
of the parameter $\lambda$ ($\lambda=100$), it has been shown elsewhere \cite{kn:12} that the equation of state obtained using the model discussed in Section 4, agrees accurately with the equation of state obtained by Horvath and Pacheco~\cite{kn:13}
for a quark-diquark mixture. Matching the static solution to the Schwarzchild
exterior solution and using the boundary condition $p_{0} = 0$ at $\bar r=b$ (note 
that in the evolutionary stage pressure does not vanish at the boundary),
we calculate the values of the constants as $R=108.779$km, $\omega = 
2.58577$ and $ D=26.6039$.

We, then, choose the constants $k_1$, $k_2$ \& $k_3$ in such a way that our model 
describes the expected early evolutionary stage of the star. 
The time dependence of $T(t)$, heat flux $q(t)$ and mass $m(r,t)$ are shown in figures 1, 2 and 3, respectively, where we considered two exemplary cases: (1) $k_{1} = 1$, $k_{2} = 10^5$ and $k_{3} = 1$ (solid lines) and  (2) $k_{1} = 1$, $k_{2} = 10^{5.2}$ and $k_{3} = 1$ (dashed lines). Arbitrary choices of these values may not give realistic results. In figures 1 and 2, we find that both $T(t)$ and $q(t)$ decreases with time, as expected. In figure 3, the mass of the star decreases with increasing time and it can be shown that as time goes to infinity the mass saturates to its final static value of $0.88~ M_{\odot}$.
 
\section{CONCLUSIONS}

Our model gives a description of the evolution of a radiating star. For a known static configuration of a star, this model generates solutions for the earlier stages of the star by a perturbative approach. This may be looked upon as complementary to the approach of Bonner {\em et al}~\cite{kn:9}. Possibly, the two methods, when combined carefully, will be able to give a total picture of the radiative collapse of a star, which ends up as a cold compact star. In our method, we have made use of the general solution given by Mukherjee {\em et al}~\cite{kn:10}. It has been shown elsewhere that the static solution can describe very compact cold stars, with an equation of state relevant for deconfined quarks \cite{kn:14} as well as a mixture of quarks and diquarks~\cite{kn:12}. The possibility that a compact star could have a deconfined quark phase in the core, surrounded by a crust of baryons and separated by a layer where the relevant density permits a hadronic phase transition, has also been considered \cite{kn:15} in the static model. Since a variety of final states are now available in the static model, the approach presented here may be useful.  

\section*{Acknowledgements}

RS acknowledges  financial support from CSIR,
New Delhi, India. TKD is thankful to the UGC, New Delhi, India, for financial
assistance and to IUCAA Reference Centre and the Physics 
Department, North
Bengal University for hospitality during the period of this work. KSG thanks the University of Natal for ongoing support.

\newpage

\begin{figure} 
\begin{center}
\vspace*{12cm}
\special{eps:hf1.eps x=13cm y=12cm}
\caption{Plot of $\frac{T(t)}{T_0}$ against $t$. ($k_{1} = 1$, $k_{2} = 10^5$ \& $k_{3} = 1$ (solid lines); $k_{1} = 1$, $k_{2} = 10^{5.2}$ \& $k_{3} = 1$ (dashed lines).)}
\end{center} \label{fig1}
\end{figure}

\newpage

\begin{figure} 
\begin{center}
\vspace*{12cm}
\special{eps:hf2.eps x=13cm y=12cm}
\caption{Variation of heat flux $q$ as a function of $t$. ($k_{1} = 1$, $k_{2} = 10^5$ \& $k_{3} = 1$ (solid lines); $k_{1} = 1$, $k_{2} = 10^{5.2}$ \& $k_{3} = 1$ (dashed lines).)}
\end{center} \label{fig2}
\end{figure}

\newpage

\begin{figure} 
\begin{center}
\vspace*{12cm}
\special{eps:hf3.eps x=14cm y=12cm}
\caption{Evolution of mass of a star whose final static configuration has mass 
$m = 0.88~M_{\odot}$. ($k_{1} = 1$, $k_{2} = 10^5$ \& $k_{3} = 1$ (solid lines); $k_{1} = 1$, $k_{2} = 10^{5.2}$ \& $k_{3} = 1$ (dashed lines). )}
\end{center} \label{fig3}
\end{figure}
\end{document}